\documentclass[%
 reprint,
superscriptaddress,
 amsmath,amssymb,
 aps,
prx,
]{revtex4-1}

\usepackage{graphicx}
\usepackage{dcolumn}
\usepackage{bm}

\begin{document}

\title{Topological-sector fluctuations and ergodicity breaking at the Berezinskii-Kosterlitz-Thouless transition}

\author{Michael F. Faulkner}
\email{michael.faulkner@bristol.ac.uk}
\affiliation{London Centre for Nanotechnology and Department of Physics and Astronomy, University College London, 17-19 Gordon Street, London WC1H 0AH, United Kingdom}
\affiliation{Laboratoire de Physique, Universit\'{e} de Lyon, \'{E}cole Normale Sup\'{e}rieure de Lyon, 46 all\'{e}e d'Italie, 69364 Lyon Cedex 07, France}

\author{Steven T. Bramwell}
\affiliation{London Centre for Nanotechnology and Department of Physics and Astronomy, University College London, 17-19 Gordon Street, London WC1H 0AH, United Kingdom}

\author{Peter C. W. Holdsworth}
\affiliation{Laboratoire de Physique, Universit\'{e} de Lyon, \'{E}cole Normale Sup\'{e}rieure de Lyon, 46 all\'{e}e d'Italie, 69364 Lyon Cedex 07, France}

\begin{abstract}

The Berezinskii-Kosterlitz-Thouless (BKT) phase transition drives the unbinding of topological defects in many two-dimensional systems. In the two-dimensional Coulomb gas, it corresponds to an insulator-conductor transition driven by charge deconfinement. 
We investigate the global topological properties of this transition, both analytically and by numerical simulation, 
using a lattice-field description of the two-dimensional Coulomb gas on a torus. 
The BKT transition is shown to be an ergodicity breaking between the topological sectors of the  
electric field, which implies a definition of topological order in terms of broken ergodicity.  The breakdown of local topological order at the BKT transition leads to the excitation of global topological defects in the electric field, corresponding to different topological sectors. The quantized nature of these classical excitations, and their strict suppression by ergodicity breaking in the low-temperature phase, afford striking global signatures of topological-sector fluctuations at the BKT transition. We discuss how these signatures could be detected in experiments on, for example, magnetic films and cold-atom systems.

\end{abstract}

\maketitle

\section{Introduction}
\label{intro}

Topological physics~\cite{Thouless} emerges in many condensed-matter systems, including 
superfluids and superconductors \cite{Onsager, Feynman,Abrikosov}, topological insulators~\cite{Hasan}, exciton-polariton condensates~\cite{Hivet}, and magnetic textures such as skyrmions~\cite{Muhlbauer}. Among two-dimensional systems, a prototypical application of topology concerns 
the quantum Hall effect in the two-dimensional electron gas~\cite{Klitzing,Geim,Kim}, while many other examples relate to the physics of topological defects identified by Berezinskii, Kosterlitz and Thouless (BKT)~\cite{Berezinskii,KTNov,Kosterlitz}. These include Josephson junction arrays~\cite{Beasley,Wolf_superconduct,Resnick,Minnhagen_superconduct_rapidPRB_1981}, films composed of Bose-Einstein condensates~\cite{Trombettoni,Hadzibabic}, superfluid films~\cite{BishopReppy}, liquid-crystal and polymer films~\cite{Birgeneau}, superinsulators~\cite{Baturina1,Baturina2}, and magnetic films and layers~\cite{BramwellHoldsworth,Willis,Elmers,Andrea}. In such systems, the BKT phase transition drives the thermal dissociation of bound pairs of local topological-defects~\cite{Berezinskii,KTNov,Kosterlitz,JKKN}. The idea of a topological defect (defined in the footnote~\footnote{An intuitive definition of a topological defect in a vector field is one that cannot be removed by continuously stretching or bending the field lines, with operations such as the discrete reversal and removal of field lines being disallowed. Under this definition, electrical charges are local topological defects in their associated electric field, as ensured by GaussÕ law. Windings of the field around the torus are global topological defects under this definition. They have no sources or sinks, yet are produced by pairs of charges tracing closed paths around the torus and annihilating each other.}) is indeed one of the most basic and important applications of topology in condensed-matter physics~\cite{Mermin}. 

An important discovery of BKT and later authors~\cite{Berezinskii,KTNov,Kosterlitz,JKKN} is that the defect-mediated transition of the plane rotator (or 2D-XY) model and its analogues can be mapped to the insulator-conductor transition of a two-dimensional Coulomb gas~\cite{Salzberg}. The long-range Coulomb interactions emerge from a purely local Hamiltonian, so that the mapping at the microscopic level, although complete~\cite{VB}, is far from transparent.
However, as Maggs and co-workers~\cite{MR,RossettoThesis,Auxiliary,LARM,MRottler,LM} have shown in three dimensions, a Coulomb fluid can be transformed into a local problem by using an electric-field representation and introducing a freely fluctuating auxiliary gauge field. 
Following this work, it is straightforward to show that the XY Hamiltonians that admit a BKT transition map on to this generalized electrostatic problem in two dimensions. A practical consequence of the phase-space extension to a fluctuating auxiliary gauge field is the development of purely local algorithms for the simulation of Coulomb fluids in both three~\cite{LARM,MR,RossettoThesis,Auxiliary,MRottler,LM} and two dimensions~\cite{Raghu}, which circumvent the technical difficulties associated with long-range interactions. 
In particular, the logarithmic potential that governs charge-charge interactions in the two-dimensional Coulomb gas is dealt with locally, allowing a new approach to the efficient simulation of two-dimensional Coulombic systems.

In this paper, we exploit these developments to formulate and simulate a lattice-field description of the two-dimensional Coulomb gas for the purpose of investigating the topological properties of the BKT transition.  The BKT transition is topological in the sense that it separates a topologically ordered phase from a disordered one. Topological order in this context means that the local topological defects (charges in the two-dimensional Coulomb gas) are confined. Vallat and Beck~\cite{VB} considered the two-dimensional XY model on a torus, and showed how a winding field can be associated with the global topology of the system. In the high-temperature phase, where charge is deconfined, non-zero values of this winding field define global topological defects that are distinct from the local topological defects driving the BKT transition. Here we show that the lattice-field description 
naturally lends itself to classifying and investigating this property.  In this paper, we treat only the two-dimensional Coulomb gas, but in a further publication we will extend our analysis to the case of two-dimensional XY models on the torus.

Our key observation is that the topology of the torus on which the Coulomb gas is placed generates a multiplicity of states in the lattice electric-field representation that are equivalent for charge configurations but not energetically degenerate. Given an arbitrary charge distribution, one is at liberty to add an integer multiple of some constant to each component of the harmonic mode of the electric field while leaving the charge distribution unchanged. This global topology associated
with the BKT transition describes the winding of charges around the torus. In the high-temperature
phase, charge deconfinement allows for fluctuations in the winding component of the harmonic mode, which can
be classified as different topological sectors. Below the transition, however, the binding of charge pairs causes
the winding component to be zero. Topological-sector fluctuations in the electric field therefore mark
the appearance of the high-temperature, topologically disordered phase at the BKT transition.

The present study of topological-sector fluctuations in the two-dimensional Coulomb gas may be compared to previous studies on the three-dimensional Coulomb phase of spin-ice materials and models~\cite{JaubertTSfluct,Moessner,JaubertKasteleyn,BrooksBartlett}. In spin ice, the onset of topological-sector fluctuations is shown to signal a Curie law crossover~\cite{JaubertTSfluct} for the zero-field susceptibility and a Kasteleyn transition in the presence of an applied field~\cite{Moessner,JaubertKasteleyn}. 
Our study of the BKT transition reveals aspects of topological-sector fluctuations that are not found in either of these established cases. For example, by our analysis, the two-dimensional Coulomb gas should be considered to present an ergodicity-breaking transition to a topologically ordered phase in the absence of an applied field, whereas spin ice has no equivalent phase.

The paper is organized as follows. In Section \ref{TSF}, we introduce the lattice-field representation of the two-dimensional Coulomb gas on a torus and use this to define the partition function and the topological sectors of the electric field. We use numerical simulations to demonstrate that
topological-sector fluctuations appear in the high-temperature (conducting) phase but not in the low-temperature (insulating) phase. In Section \ref{ergodicity}, we show that the reason for the strict suppression of topological-sector fluctuations in the 
low-temperature phase is ergodicity breaking at the transition. A finite-size scaling analysis, given in Section \ref{FSS}, confirms that in the thermodynamic limit, ergodicity is broken precisely at $T_{\rm BKT}$. Conclusions and comparisons with experimental systems are discussed in Section \ref{conclusions}.

\section{Topological-sector fluctuations }\label{TSF}
\label{TS}

Using the unit system defined in Appendix \ref{units}, we formulate the two-dimensional Coulomb gas using discrete vector calculus on a square lattice with periodic boundary conditions (PBCs) applied. The PBCs enforce the toroidal topology but not the curvature of a true torus. All functions are defined to be the discrete counterparts of smooth vector fields~\cite{Chew}, and any lattice vector field $\mathbf{F}$ is defined component-wise via~\cite{Chew}
\begin{align}
\mathbf{F}(\mathbf{x}):=F_x\left( \mathbf{x}+\frac{a}{2}\mathbf{e}_x\right) \, \mathbf{e}_x + F_y\left( \mathbf{x}+\frac{a}{2}\mathbf{e}_y\right) \, \mathbf{e}_y,
\end{align}
where $\mathbf{x}$ is any lattice site and $\mathbf{e}_{x/y}$ is the unit vector in the $x/y$ direction. The operators $\boldsymbol{\tilde{\nabla}}$ and $\boldsymbol{\hat{\nabla}}$ are the forwards and backwards finite-difference operators on a lattice, respectively, and the lattice Laplacian is defined by $\boldsymbol{\nabla}^2:=\boldsymbol{\hat{\nabla}} \cdot \boldsymbol{\tilde{\nabla}}$~\cite{Chew}. The most general electric field $\mathbf{E}$ 
may be Helmholtz decomposed into the sum of a Poisson (divergence-full) component $-\boldsymbol{\tilde{\nabla}}\phi$, a rotational component 
$\mathbf{\tilde{E}}$ and a harmonic component $\mathbf{\bar{E}}$:
\begin{align}\label{field}
\mathbf{E}(\mathbf{x})=-\boldsymbol{\tilde{\nabla}}\phi(\mathbf{x})+\mathbf{\tilde{E}}(\mathbf{x})+\mathbf{\bar{E}}. 
\end{align}
This electric field is the most general solution to Gauss' law on a lattice: 
\begin{align}
\label{gauss}
\boldsymbol{\hat{\nabla}} \cdot \mathbf{E}(\mathbf{x})=\rho(\mathbf{x})/ \epsilon_0.
\end{align}
Here, $\rho(\mathbf{x}):=qm(\mathbf{x})/a^2$ is the charge density at each lattice site $\mathbf{x}$,  $q$ is the elementary charge, the integer $m$ denotes the charge species, $a$ is the lattice spacing and $\epsilon_0$ is the electric permittivity of free space (see Appendix \ref{units}). 
Using the field $\mathbf{E}$ adds an auxiliary field $\mathbf{\tilde{E}}$ to the usual solution of electrostatics, as in the electrostatic model of Maggs and Rossetto (MR)~\cite{MR}.  This allows us to simulate the physics of Coulombic interactions on a lattice via local electric-field updates, avoiding the need to treat computationally intensive long-range interactions, as outlined in Appendix \ref{appPF}.  

The validity of introducing the auxiliary field is seen in the context of the separability of the partition function into its Coulombic and auxiliary components: the auxiliary field contributes to the internal energy of the system, but it is statistically independent of the Coulombic element. In Appendix \ref{appPF}, we give a full description of the algorithm and a derivation of the partition function for the Coulomb gas of multi-valued charges.

The internal energy of the electric fields corresponding to a given charge and auxiliary-field configuration is given by
\begin{align}\label{u0}
U_0 = \frac{\epsilon_0 a^2}{2}\sum_{\mathbf{x}\in D} |\mathbf{E}(\mathbf{x})|^2,
\end{align}
where 
$D$ is the set of all lattice points. To represent the Coulomb gas in the grand canonical ensemble, we add a core-energy term $U_{\text{Core}}$, given by
\begin{align}\label{core}
U_{\text{Core}}:=\frac{a^4}{2}\sum_{\mathbf{x}\in D}\epsilon_c \left[ m(\mathbf{x})\right] \rho(\mathbf{x})^2,
\end{align}
where 
$\epsilon_c(m)$ is the core-energy constant of each charge $mq$, and 
$\epsilon_c(m)=\epsilon_c(-m)$ since charges are excited to the vacuum in neutral pairs. 
The grand-canonical energy of the system $U = U_0 + U_{\rm Core}$ may be expanded by combining Eqs. (\ref{field}), (\ref{u0}) and  (\ref{core}) 
to give a sum of terms arising from the different field components, which add to the core energy:
\begin{align}
U=U_{\text{Self}}+U_{\text{Int}}+U_{\text{Rot}}+U_{\text{Harm}}+U_{\text{Core}}.
\end{align}
Here, respectively, $U_{\text{Rot}} := \epsilon_0 a^2\sum_{\mathbf{x}\in D} |\mathbf{\tilde{E}}(\mathbf{x})|^2/2$ and $U_{\text{Harm}} := \epsilon_0 L^2 |\mathbf{\bar{E}}|^2/2$
are the auxiliary-field and harmonic-mode components of the grand-canonical energy, and $U_{\text{Self}}$ and $U_{\text{Int}}$ are the self-energy and Coulombic charge-charge interaction 
components. As outlined in detail in Appendix \ref{appPF}, the latter two components may be expressed in terms of the lattice Green's function
$G$, according to 
 $U_{\text{Self}}:=a^4\, G(\mathbf{0})\sum_{\mathbf{x}\in D}\rho(\mathbf{x})^2/2\epsilon_0$, $U_{\text{Int}}:=a^4\sum_{\mathbf{x}_i\ne \mathbf{x}_j\in D}\rho(\mathbf{x}_i)G(\mathbf{x}_i,\mathbf{x}_j)\rho(\mathbf{x}_j)/2\epsilon_0$, where $G(\mathbf{0}):=G(\mathbf{x},\mathbf{x})$. Note that, while 
 $U_{\rm Int}$ can be negative, the 
 sum $U_{\text{Self}}+U_{\text{Int}}$ is necessarily $\ge 0$, as it arises from the term in $|\boldsymbol{\tilde{\nabla}} \phi |^2$.
 
Using the above results, we may define the chemical potential for the introduction of a charge $ mq$:
\begin{align}
\mu_m:=-\left[ \frac{G(\mathbf{0})}{\epsilon_0}+\epsilon_c(m) \right] \frac{m^2q^2}{2}.
\end{align}
In the following, we specialize to a Coulomb gas of  $n$ pairs of elementary charges of chemical potential $\mu :=\mu_1$, by setting
 $\epsilon_c(m= 0,\pm 1)=0$ and $\epsilon_c(m\ne 0,\pm 1)=\infty$ 
~\footnote{Note that the BKT transition is not restricted to a system of elementary charges: the charges can be multi-valued, as in Villain's model: J. Villain, J. Physique, {\bf 36}, 581 (1975)}. 

The harmonic mode $\mathbf{\bar{E}}$ is a uniform field found by averaging the total electric field $\mathbf{E}(\mathbf{x})$ over  ${\bf x}$. In a simply connected space, the average field 
may be conveniently related to the average polarization $\mathbf{P}$ arising from an effective surface charge distribution by $\mathbf{\bar{E}} = -\mathbf{P}/\epsilon_0$. For a charge-neutral system in a simply connected space, $\mathbf{P}:= \sum_{\mathbf{x}\in D} \mathbf{x} \rho(\mathbf{x})/N$ is invariant with respect to the origin shift $\mathbf{x} \mapsto \mathbf{x}+\mathbf{x}_0$, and is therefore origin-independent. The situation is more complicated on a toroidal surface as $\mathbf{\bar{E}}$
can also depend on a harmonic-field component that corresponds to a charge winding around the torus, and it is necessary to adopt a convention to define distances between points (the concepts `close together' and `far apart' are ambiguous on a  torus). In Appendix \ref{Polarization}, we show in detail how it is possible to define origin-independent polarization
$\mathbf{\bar{E}}_{\textrm{p}}$ and winding  $\mathbf{\bar{E}}_{\textrm{w}}$ components of the harmonic mode such that
\begin{align}\label{approach2}
\mathbf{\bar{E}} =\mathbf{\bar{E}}_{\mathrm{p}}+\mathbf{\bar{E}}_{\mathrm{w}}.
\end{align}
Here,
\begin{align}\label{Ebarw}
\mathbf{\bar{E}}_{\mathrm{w}}=\frac{q}{L\epsilon_0}\mathbf{w},
\end{align}
where $\mathbf{w}$ is an integer-valued winding field chosen such that 
\begin{align}
\label{dipset}
\bar{E}_{\mathrm{p},x/y}\in \left(-\frac{q}{2L\epsilon_0},\, \frac{q}{2L\epsilon_0}\right],
\end{align}
and $L$ is the lattice length. 

This decomposition of the harmonic field has the following interpretation. A charge pair may unbind and wind around the torus in opposing directions before assuming its original configuration.  When a single charge winds around the torus in the $x/y$ direction, 
the $x/y$ component of the harmonic mode of the electric field $\bar{E}_{x/y}$ increases by $\pm q/L\epsilon_0$. As shown in Appendix \ref{Polarization}, the lowest-energy harmonic mode that describes an arbitrary charge distribution is therefore an element of the set $\left(-q/2L\epsilon_0 ,\, q/2L\epsilon_0 \right]$ 
and is defined as the polarization component in Eq. (\ref{dipset}) by applying modular arithmetic to $\mathbf{\bar{E}}$.  The remainder is the winding component. The modulo operation removes any need for a `distances' convention to define the polarization component, as well as any origin dependence of the field components (see Appendix \ref{Polarization} for further details).

With these results we may use the integer-valued winding field $\mathbf{w}$ to define the topological sector of the system as the number of times charges wind around the torus in the $x$ and $y$ directions, with all non-trivial topological sectors given by $\mathbf{w}\ne \mathbf{0}$. The topological sector of the system changes any time a charge pair unbinds and winds around the torus and hence thermal fluctuations of the topological sector are closely related to the unbinding of charge pairs, as elucidated further below.   

The statistical mechanics of the topological-sector fluctuations may now be formulated by considering how the polarization and winding components of the harmonic mode enter the lattice partition function. As shown in Appendix B, the partition function splits into two statistically independent components. One component is the Coulombic partition function $Z_{\text{Coul}}$ and contains all information about the charge-charge correlations, while the other is the auxiliary-field partition function $Z_{\text{Rot}}$ and contains all information about the auxiliary field: the auxiliary field can freely fluctuate without affecting charge-charge correlations (see Appendix \ref{appPF}). The partition function is written as
\begin{align}
Z=Z_{\text{Coul}}Z_{\text{Rot}},
\end{align}
where $Z_{\text{Coul}}$ is given by
\begin{align}
\label{PFfinal}
Z_{\text{Coul}}=&\sum_{\{ \rho(\mathbf{x}) \} }\exp \left[ -\frac{\beta a^4}{2\epsilon_0}\sum_{\mathbf{x}_i\ne \mathbf{x}_j}\rho(\mathbf{x}_i)G(\mathbf{x}_i,\mathbf{x}_j)\rho(\mathbf{x}_j) \right] \nonumber\\
&\times \sum_{\mathbf{w}\in \mathbb{Z}^2}\exp\left( -\frac{L^2\beta \epsilon_0}{2}|\mathbf{\bar{E}}_{\mathrm{p}}+\frac{q}{L\epsilon_0}\mathbf{w}|^2 \right) \nonumber\\
&\times \delta \left( \sum_{{\bf x}\in D}\rho ({\bf x}) \right) e^{\beta \mu n} .
\end{align}
Here, $\beta:=1/k_{\textrm{B}}T$ is the inverse temperature and the sum $\sum_{\{ \rho(\mathbf{x}) \} }:= \sum_{\{ a^2\rho(\mathbf{x})\in \{ 0,\pm q\} \} }$.

\begin{figure}
\includegraphics[width=\linewidth]{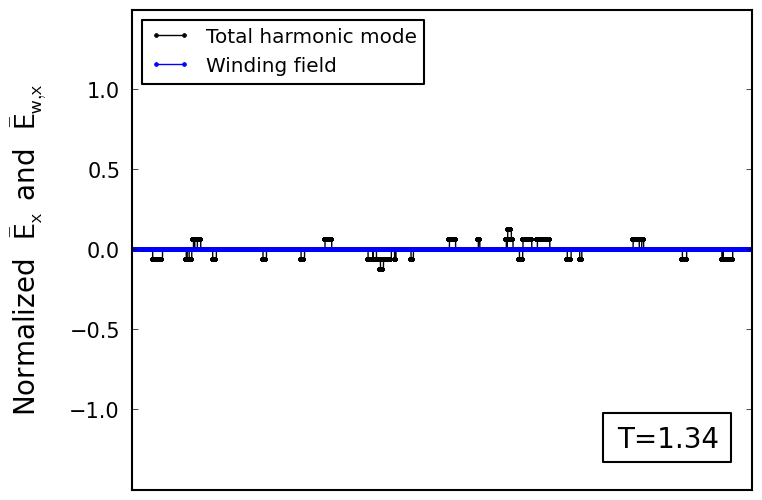}
\includegraphics[width=\linewidth]{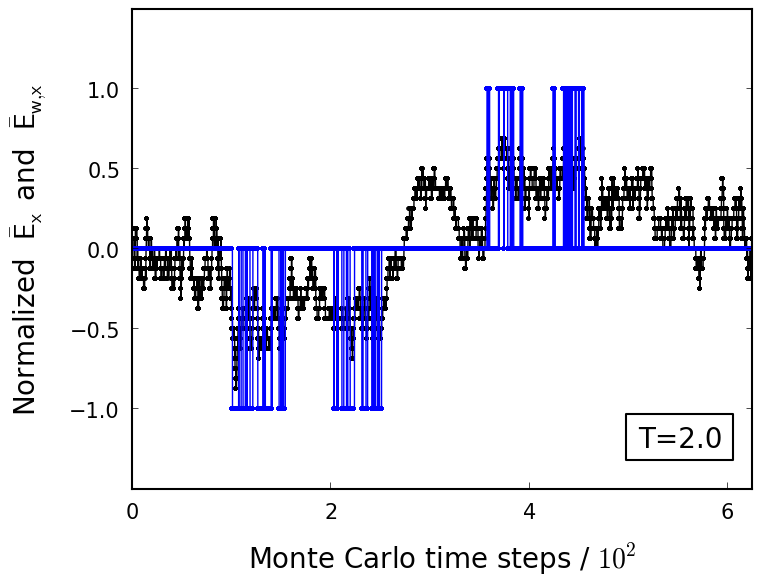}
\caption{Topological-sector fluctuations of lattice electric fields in the two-dimensional Coulomb gas on a torus. Shown is the $x$-component of the normalized total harmonic mode $L \bar{E}_x/2\pi$ and winding field $L \bar{E}_{\textrm{w},x}/2\pi$ 
versus Monte Carlo time for an $L\times L$ system of linear size $L=16$ at $T=1.34$ (top) and $T=2.0$ (bottom). The system was simulated using the MR algorithm with local moves only. At the lower temperature (top) harmonic-mode fluctuations are finite (black) but there is no winding-field component (blue), while at the higher temperature (bottom) the winding-field component becomes finite, indicating topological-sector fluctuations. }\label{dynamic}
\end{figure}

The first exponential of Eq. (\ref{PFfinal}) describes the anharmonic charge-charge interactions, the second describes the polarization and winding state of the system, and the third describes the sum of the self-energies associated with each elementary charge. The sum over the winding field $\mathbf{w}$ is necessitated by the degeneracy of the harmonic mode of the electric field: infinitely many topological sectors describe any given charge configuration. In addition to the local updates of the MR algorithm (see Appendix \ref{appPF}), we also consider global updates,  which correspond to independently sampling this winding field.

Henceforth, we set the elementary charge $q=2\pi$, the lattice spacing $a=1$, the electric permittivity $\epsilon_0 =1$, and Boltzmann's constant $k_{\textrm{B}}=1$. The choice $q=2\pi$ recognizes the standard BKT theory, where a charge emerges as a local $2\pi$ winding in an associated lattice field, such as the spin differences in the 2D-XY model~\cite{JKKN}.

The BKT transition drives the deconfinement of charge pairs in the two-dimensional lattice Coulomb gas, which generates topological-sector fluctuations. The transition occurs at $T_{\rm BKT}=1.351$ (to four significant figures)~\cite{Janke} in the thermodynamic limit [a value specific to a gas of elementary charges with $\epsilon_c(m=1)=0$], which is scaled to higher temperatures in finite-size systems (see below). Fig. \ref{dynamic} shows the evolution of the (normalized) $x$-component of the harmonic mode of a system of linear size $L=16$, simulated using local moves only (numerical simulation details are described in Appendix \ref{sim_details}). No topological-sector fluctuations are visible just below the BKT transition temperature $T_{\rm BKT}=1.351$, but they become important at temperatures above $T_{\rm BKT}$. 

\section{Ergodicity breaking}\label{ergodicity}

A convenient measure of the topological-sector fluctuations is the winding-field susceptibility $\chi_{\textrm{w}}$, defined by
\begin{align}\label{def_wind_susc}
\chi_{\textrm{w}}(L,T):=\beta \epsilon_0 L^2 \left( \langle \mathbf{\bar{E}}_{\textrm{w}}^2 \rangle -\langle \mathbf{\bar{E}}_{\textrm{w}} \rangle^2 \right).
\end{align}
In a fully ergodic system, $\chi_{\textrm{w}}$ is nonzero, even in the absence of charge fluctuations, as can be seen by limiting the Gibbs ensemble contributing to $Z_{\text{Coul}}$ to field configurations of zero charge. In this case it is straightforward to show, using Eqs. (\ref{Ebarw}) and (\ref{PFfinal}), that the constrained susceptibility is given by
\begin{align}\label{w-global}
\chi_{\textrm{w}}^{\textrm{global}}(T)&=\beta \epsilon_0 L^2\frac{4q^2 \exp \left( -\beta q^2/2\epsilon_0 \right)/\epsilon_0^2 L^2 +\dots}{1+4\exp \left( -\beta q^2/2\epsilon_0 \right) +\dots}\nonumber\\
&\simeq \frac{4\beta q^2}{\epsilon_0} \exp \left( -\beta q^2/2\epsilon_0 \right) ,
\end{align}
for $k_{\textrm{B}}T\ll q^2/2\epsilon_0$. The system-size dependence falls out of this expression 
so that a fully ergodic system would show small but finite topological-sector fluctuations in the low-temperature phase.

Assuming local charge dynamics, a topological-sector fluctuation requires the separation of a pair of charges 
over a distance greater than $L/2$ in either the $x$ or the $y$ direction [see Eq. (\ref{approach2}) and the subsequent discussion]. As the charge concentration falls to zero at low temperature, screening becomes negligible and the energy barrier against such configurations diverges logarithmically with the linear system size $L$~\cite{Salzberg,Berezinskii,KTNov}. As the charge concentration increases with temperature, however, entropy and charge screening break down the free-energy barrier, making it finite at the BKT transition. Above the transition, charge pairs are free to unbind and trace closed paths around the torus, giving finite-valued winding fields, as observed in Fig. \ref{dynamic}. In contrast, in the low-temperature phase, 
the probability of separation through a distance $L/2$ becomes strictly zero in the thermodynamic limit.

The BKT transition is therefore an ergodicity breaking: a change in the phase space explored by a system with local dynamics. In detail, it is an ergodicity breaking between topological sectors, signalled by the strict suppression of topological-sector fluctuations in the electric field at $T < T_{\rm BKT}$. If the dynamics were non-local (including global updates of the winding component of the harmonic mode~\cite{LARM,MR,RossettoThesis,Auxiliary,MRottler,LM}), $\chi_{\textrm{w}}$ would remain finite at all temperatures.

To explore this ergodicity breaking, we have simulated the two-dimensional Coulomb gas, first with local field updates only, and second with both local and global field updates~\cite{LARM,MR,RossettoThesis,Auxiliary,MRottler,LM}. Corresponding to each case, we define the winding-field susceptibilities $\chi_{\textrm{w}}^{\text{local}}$ and $\chi_{\textrm{w}}^{\text{all}}$, respectively. Differences between $\chi_{\textrm{w}}^{\text{local}}$ and $\chi_{\textrm{w}}^{\text{all}}$ reflect the inability of local moves to explore a fully representative phase space on the time scale of the simulation. To quantify this, we introduce the susceptibility quotient $\chi_{\textrm{w}}^{\text{local}}/\chi_{\textrm{w}}^{\text{all}}$, which may be used to analyse the ergodicity of the system. 

\begin{figure}
\includegraphics[width=\linewidth]{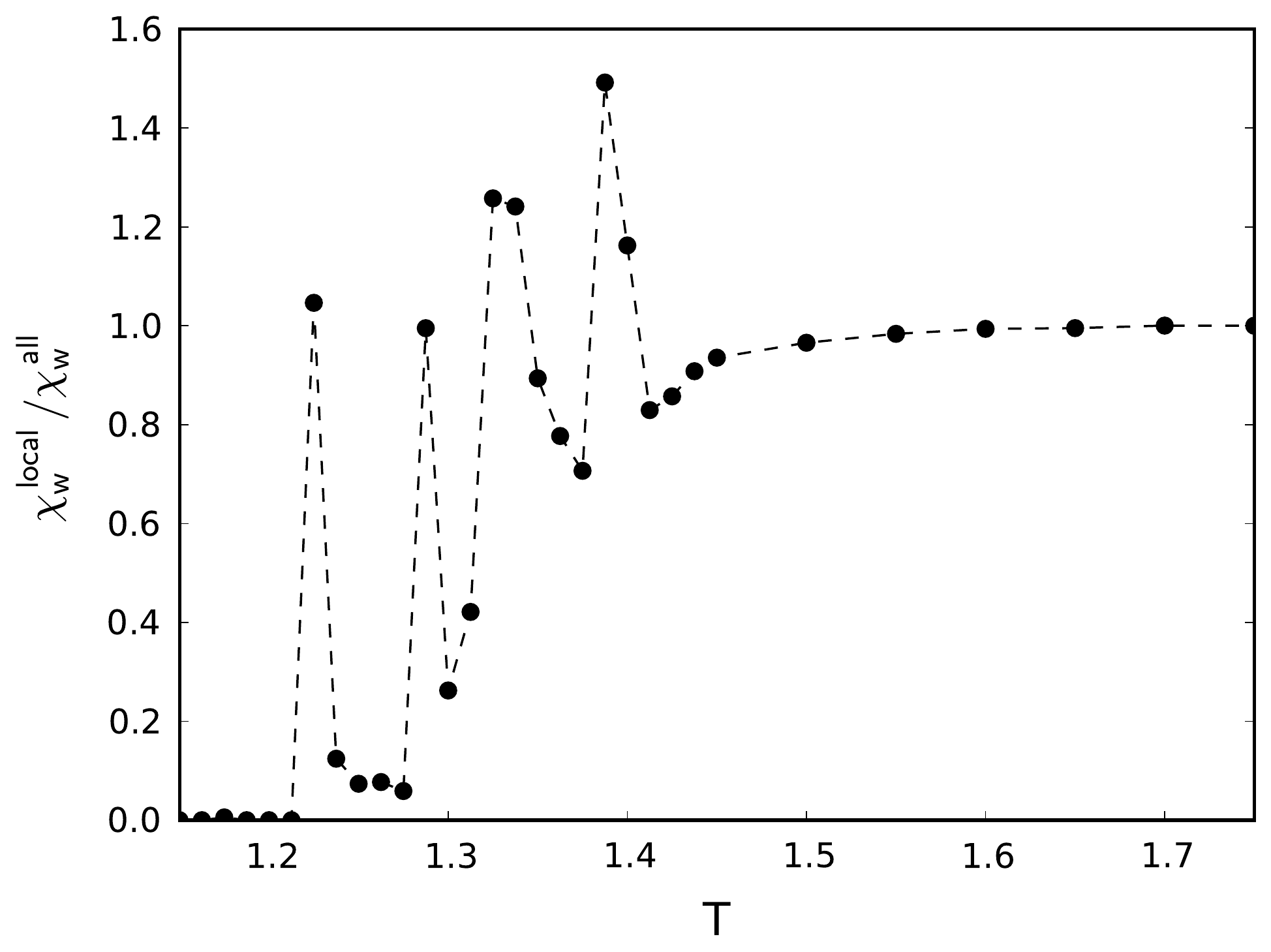}
\caption{The susceptibility quotient $\chi_{\textrm{w}}^{\text{local}} / \chi_{\textrm{w}}^{\text{all}}$ versus temperature for an $L\times L$ Coulomb gas of linear size $L=64$. In the region $T<1.2$, the quotient is zero, while for $T>1.6$, the quotient approaches unity. This divergence between the results of the local-update and the all-updates simulations, accompanied by striking fluctuations in the intermediate region, signals an ergodicity breaking as the system is cooled through the BKT transition. The line is a guide to the eye. } \label{figsuscquotient}
\end{figure}

Fig. \ref{figsuscquotient} clearly shows that ergodicity is broken in the vicinity of the BKT transition. For $T>1.6$,
$\chi_{\textrm{w}}^{\text{local}} = \chi_{\textrm{w}}^{\text{all}}$, indicating that the free-energy barrier for a topological-sector fluctuation via local moves is small. For $T<1.2$, 
the quotient is zero, indicating that the energy barrier prevents topological-sector fluctuations via local charge moves. In between these low- and high-temperature regions  there are strong
fluctuations in the quotient because charge deconfinement via local updates represents increasingly rare events, an inevitable precursor to loss of ergodicity. In Section \ref{FSS}, this ergodicity breaking is shown to occur precisely at $T_{\textrm{BKT}}$ in the thermodynamic limit.

Our analysis thus leads to a precise definition of topological order for the two-dimensional Coulomb gas through the ergodic freezing of the topological sector to its lowest absolute value. Two-dimensional systems with $U(1)$ symmetry are often associated with an absence of an ordering field at finite temperature~\cite{MerminWagner}. Here we explicitly show that, in the case of the BKT transition, the ordering of a conventional order parameter is replaced by topological ordering through an ergodicity breaking between the topological sectors.  The topological order is directly related to the confinement-deconfinement transition of the charges, the local topological defects of the electric field. This type of ergodicity breaking is distinct from either the symmetry breaking that characterizes a standard phase transition, or that due to the rough free-energy landscape that develops at a spin-glass transition~\cite{Palmer1982}.

\section{Finite-size scaling}\label{FSS}

In order to explore the approach to the thermodynamic limit, the two-dimensional Coulomb gas was simulated by the Monte Carlo method 
as a function of system size, using the MR algorithm. The global update was employed in order to improve the statistics
(numerical simulation details are described in Appendix \ref{sim_details}).

Fig. \ref{figfullsusc} shows the simulated winding-field susceptibility $\chi_{\textrm{w}}$ as a function of temperature for $L\times L$ Coulomb gases of linear sizes between
 $L=8 $ and $L=64$. There is a marked increase in the winding-field susceptibility $\chi_{\mathrm{w}}$ as the system passes through the BKT transition temperature $T_{\rm BKT}=1.351$~\cite{Janke} for all system sizes. Susceptibility curves for successive values of $L$ intersect at temperatures above $T = 1.8$ and below $T = 1.5$. Between these two temperatures, the winding-field susceptibility increases for a given temperature as the linear system size $L$ increases. 
These results  are consistent with the finite-size scaling of the BKT transition temperature~\cite{BramwellHoldsworth}: as the system size decreases the effective transition temperature $T^*(L)$ increases.

\begin{figure}[]
\includegraphics[width=\linewidth]{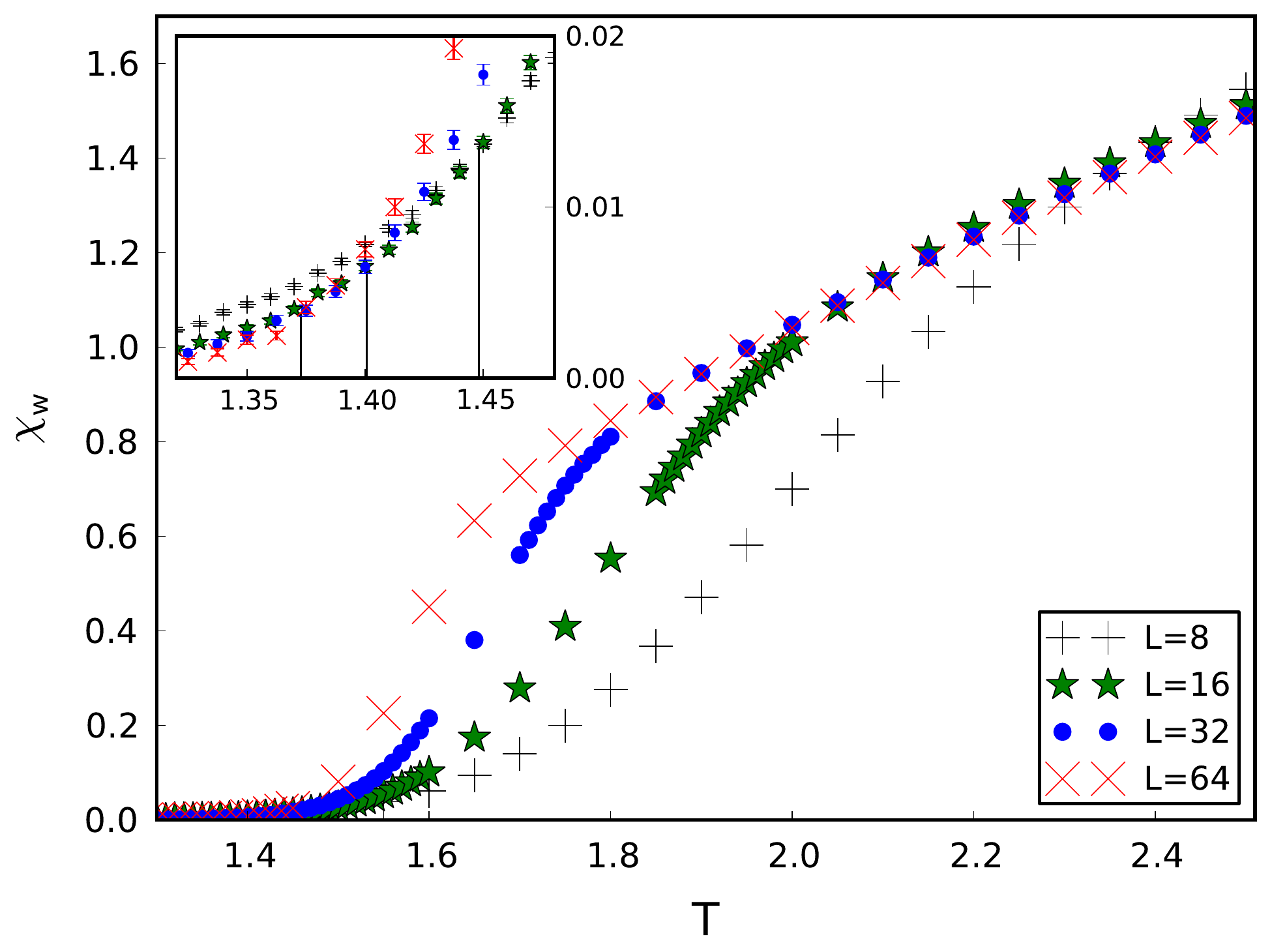}
\caption{The winding-field susceptibility $\chi_{\textrm{w}}$ as a function of temperature for $L\times L$ Coulomb gases of linear size $L=8,\, 16,\, 32$ and $64$ (using local and global MR moves). The curves intersect at low and high temperature. Inset: An expanded plot of the data in the region of the low-temperature intersections (with error bars representing two standard deviations). 
The indicated crossover temperatures are given by $T_{\text{Cross}}(L=16)=1.45$, $T_{\text{Cross}}(L=32)=1.40$ and $T_{\text{Cross}}(L=64)=1.37$ (to within estimated error), based on a data fit.}
\label{figfullsusc}
\end{figure}
Below $T_{\textrm{BKT}}$, the probability of a charge pair separating over a distance greater than $L/2$ increases with decreasing system size. This, combined with the finite-size transition temperature $T^*(L)$ also increasing with decreasing system size, results in the winding-field susceptibility curves for successive values of $L$ intersecting in the vicinity of $T_{\textrm{BKT}}$. The inset in Fig. \ref{figfullsusc} shows that the low-temperature crossover points of the susceptibility curves are at $T=1.45$, $T=1.40$ and $T=1.37$ (to within estimated error). To extrapolate the trend of the data shown in Fig. \ref{figfullsusc} to the thermodynamic limit, we define the crossover temperature $T_{\text{Cross}}(L)$ to be the lower temperature at which $\chi_{\textrm{w}}(L)=\chi_{\textrm{w}}(L/2)$. 

\begin{figure}[]
\includegraphics[width=\linewidth]{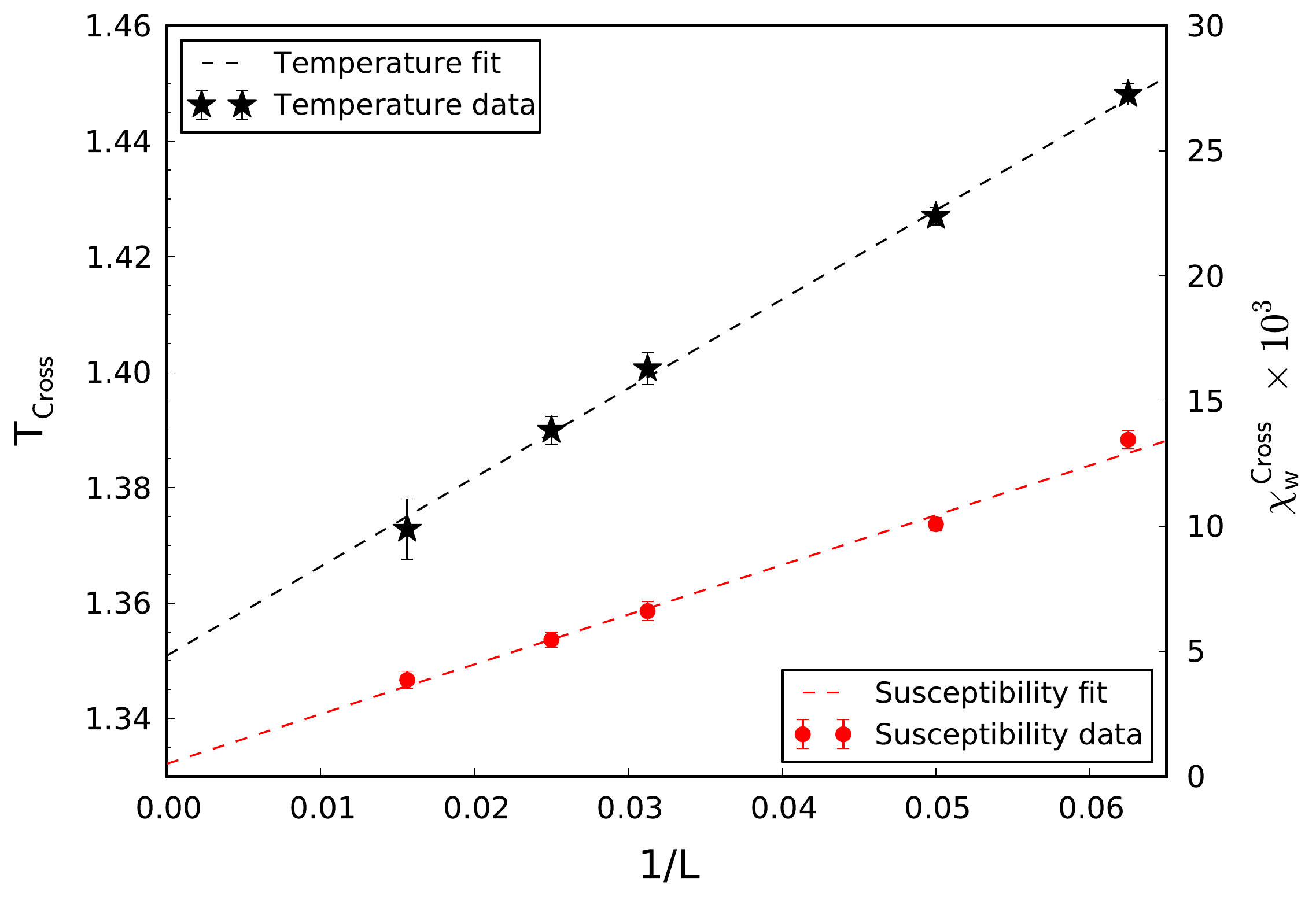}
\caption{The crossover temperature $T_{\text{Cross}}$ (black data; left-hand $y$ axis) and crossover susceptibility $\chi_{\mathrm{w}}^{\mathrm{Cross}}$ (red data; right-hand $y$ axis) as functions of inverse linear system size $1/L$, with error bars representing two standard deviations. Lines are weighted (with respect to the error bars) linear-regression fits to each data set, from which the $y$-intercept ($L\rightarrow \infty$) was calculated. $T_{\textrm{Cross}}(L\rightarrow \infty)=1.351(2)$, equal to the BKT transition temperature $T_{\rm BKT}$~\cite{Janke}. 
The crossover susceptibility $\chi_{\textrm{w}}^{\textrm{Cross}}(L\rightarrow \infty)=5\times 10^{-4}$ with estimated error of 
the same order: 
there is no measurable difference between this quantity and the winding-field susceptibility due to global updates only at $T=1.351$.} \label{figtempextrap}
\end{figure}

Fig. \ref{figtempextrap} shows the crossover temperature $T_{\mathrm{Cross}}$ as a function of inverse linear system size $1/L$, along with straight-line fits to the data. 
In the thermodynamic limit, $T_{\text{Cross}}$ extrapolates to the value  $T_{\text{Cross}} = 1.351$ to within the estimated error of the extrapolation, that is, it extrapolates to the BKT transition temperature \cite{Janke}:
\begin{align}
T_{\text{Cross}}(L\rightarrow \infty)=T_{\rm BKT}.
\end{align}
The $1/L$ scaling of $T_{\text{Cross}}$ is unusual for the BKT transition, for which the finite-size BKT transition temperature typically scales as a simple function of the logarithm of $L$~\cite{BramwellHoldsworth,Weber_MinnhagenPRB}. However, Minnhagen and Kim~\cite{Minnhagen_KimPRB} found that a fourth-order cumulant that measures fluctuations of the helicity modulus in the 2D-XY model also scales as $1/L$: as this closely relates to fluctuations in the harmonic-mode susceptibility~\cite{VB}, it seems likely that we are observing the same finite-size scaling here. The magnitude of the winding-field susceptibility at the crossover points $\chi_{\mathrm{w}}^{\mathrm{Cross}}(L\rightarrow \infty)$ similarly extrapolates 
to $\sim 5\times 10^{-4}$ in the thermodynamic limit, with an estimated error of the same order. This small number is not measurably different to the winding-field susceptibility due to global moves only, which, at $T_{\textrm{BKT}}$, evaluates to $\sim 5\times 10^{-5}$ for all system sizes [see Eq. (\ref{w-global})]. The inference is that topological-sector fluctuations due to local moves only turn on precisely at the universal point $T_{\text{Cross}}(L\rightarrow \infty)=T_{\rm BKT}$ in the thermodynamic limit. This confirms that topological-sector fluctuations signal charge 
deconfinement and the high-temperature phase of the BKT transition.

Given that the topological-sector fluctuations turn on at the temperature at which the system experiences the famous universal jump in the helicity modulus~\cite{Kosterlitz,JKKN,VB,Minnhagen_KimPRB}, it is interesting to estimate the contribution that topological-sector fluctuations make to the universal jump. 
To do this, we define the harmonic-mode susceptibility $\chi_{\mathbf{\bar{E}}}$ and the polarization susceptiblity $\chi_{\textrm{p}}$ by replacing $\mathbf{\bar{E}}_{\textrm{w}}$ in Eq. (\ref{def_wind_susc}) with $\mathbf{\bar{E}}$ and $\mathbf{\bar{E}}_{\textrm{p}}$, respectively. The helicity modulus is then given by 
$\Upsilon = \epsilon_0^{-1}\left( 1-\chi_{\bar{\mathbf{E}}}/2\right)$~\cite{VB}, so that $\chi_{\bar{\bf E}}$ makes a jump of order unity at $T_{\rm BKT}$. We find that the ratio $(\chi_{\mathbf{\bar{E}}}-\chi_{\textrm{p}})/\chi_{\mathbf{\bar{E}}}$ is less than $5\times 10^{-2}$ for all $T\le 1.6$ for systems of linear size $L=8$ to $64$, showing that the contribution to the universal jump from topological-sector fluctuations is very small. This reflects the near-cancellation of 
$\langle \mathbf{\bar{E}}_{\textrm{w}}^2\rangle$ 
with the coupling term $2\langle \mathbf{\bar{E}}_{\textrm{p}}\cdot \mathbf{\bar{E}}_{\textrm{w}}\rangle$ in the evaluation of $\chi_{\bar{\bf E}}$, reflecting strong correlations
between the polarization and winding fields at the transition.

\section{Conclusions}
\label{conclusions}

In conclusion, the BKT transition has long been a paradigm for the importance of topological defects in condensed-matter physics~\cite{Thouless}. 
Vallat and Beck showed that XY-type systems on the torus generate global topological defects at the BKT transition that reflect the toroidal topology~\cite{VB}. 
Here we have used lattice-field simulations to reveal topological-sector fluctuations in the electric field of a two-dimensional lattice Coulomb gas on a torus. 
We have shown how these provide a striking and sensitive measure of the topological and ergodicity-breaking character of the BKT transition,
allowing a precise definition of topological order in terms of this broken ergodicity. 

The topological-sector fluctuations at the BKT transition are very clearly revealed in the lattice electric field description of the two-dimensional Coulomb gas, but we expect them to be equally relevant to any system that has a BKT transition. In suitable systems, the winding-field susceptibility that signals the onset of topological-sector fluctuations will contribute to experimentally measurable responses of the system. For example, in a cylindrical or toroidal magnetic film with XY symmetry, 
winding-field fluctuations in the Coulomb gas representation correspond to measurable spin configurations in the magnetic representation. 
As we will show in future work~\cite{FBH_MRXY}, fluctuations of an appropriate topological sector accompany the destruction of the finite-size magnetization of an XY spin system through vortex deconfinement. They could therefore be observable in ultrathin ferromagnetic metallic films~\cite{Ahlberg} or magnetic Langmuir-Blodgett films~\cite{Mukhopadhyay,Gayen}. 

Another promising system on which to measure these topological-sector fluctuations is the one-dimensional quantum lattice Bose gas. 
When the system is placed on a ring, its angular momentum is no longer a good quantum number.
The angular momentum can therefore fluctuate quantum mechanically, and the system should undergo a dramatic increase in these fluctuations as it passes through the superfluid -- Mott insulator quantum phase transition~\cite{Greiner,Stoferle}. This dramatic increase in the fluctuations corresponds to finite-valued global topological defects in the quantum system, and therefore, via the Feynman path-integral mapping, to topological-sector fluctuations in the two-dimensional classical lattice Coulomb gas on a torus. Murray {\it et al.} measured the angular momentum of 
ring-shaped Bose-Einstein condensates via the vortex-density profile of the system~\cite{Murray}. Our measure of the BKT transition could therefore correspond to equivalent, experimentally measurable topological-sector fluctuations in the cold-atom system~\cite{Roscilde2016FromQuantum}.

Finally, it is worth noting that it is natural to associate a conducting phase with the excitation of winding fields, as may be seen by considering a loop of wire in a changing magnetic field. Recalling that the magnetic field does no work on a test charge, the induced electromotive force must arise from a divergence-free electric field running round the loop. The curl of this field obeys the Maxwell-Faraday law, $\boldsymbol{\nabla} \times {\bf E} = \partial {\bf B}/\partial t$.  Hence, in three dimensions, electromagnetic induction provides a practical method of exciting topological winding fields analogous to those discussed here. 

\begin{acknowledgements}
It is a pleasure to thank A. C. Maggs, S. T. Banks, V. Kaiser and G. B. Davies for valuable discussions, and A. Gormanly for help with automating repeated simulations. We are also grateful to T. Roscilde for pointing out the possible application to the one-dimensional lattice Bose gas. M.F.F. is grateful for financial support from the CNRS and University College London. P.C.W.H. acknowledges financial support from the Institut Universitaire de France.
\end{acknowledgements}

\appendix

\section{Dimensional analysis of the two-dimensional Coulomb gas}
\label{units}
In the following,  $[\, \cdot \, ]$ denotes the dimensions of some quantity, $L$ denotes the dimensions of length, $d$ is the spatial dimensionality of the system, and $\epsilon_0$ is the vacuum permittivity in $d-$dimensional space.  
Consider Gauss' law for the MR algorithm,
\begin{align}
\boldsymbol{\hat{\nabla}}\cdot \mathbf{E}(\mathbf{x})=\rho(\mathbf{x})/ \epsilon_0 ,
\end{align}
and the dimensions of electric charge density,
\begin{align}
\left[ \rho(\mathbf{x}) \right] = [q]\, L^{-d},
\end{align}
which generates
\begin{align}
\left[ \mathbf{E}(\mathbf{x}) \right] = [q]\, L^{(1-d)}\left[ \epsilon_0 \right]^{-1}. 
\end{align}
\noindent 
From a consideration of the exponent of the Boltzmann factor (with $\beta = 1/k_{\rm B}T$) we find a dimensionless group,
\begin{align}
\Pi =& \frac{a^d\beta \epsilon_0}{2} 
\sum_{\mathbf{x} \in D} |\mathbf{E}(\mathbf{x})|^2\\
\Rightarrow \left[ \epsilon_0 \right] =& [q]^2 L^{(2-d)} \left[ \beta \right] \\
\Rightarrow \left[ \mathbf{E}(\mathbf{x}) \right] =& [q]^{-1} L^{-1} \left[ \beta \right]^{-1} . 
\end{align}
Setting the charge to be dimensionless, it follows that 
\begin{align}
\left[ \epsilon_0 \right] = \left[\beta \right]
\end{align}
and
\begin{align}
\left[ \mathbf{E}(\mathbf{x}) \right] = [\beta]^{-1} L^{-1}
\end{align}
in $d=2$.

\section{The MR electrostatic model and the partition function}\label{appPF}

The MR electrostatic model is a lattice-field model from which it is possible to form the lattice partition function of electrostatics. To show this, we describe the MR algorithm in terms of microscopic variables that represent the local field updates. A conjugate lattice $D'$ is defined such that each of its sites is at the centre of each plaquette of $D$. Each site in $D'$ is associated with a real-valued variable $\varphi$ whose adjustment corresponds to an update of the auxiliary field, while each pair of nearest-neighbour sites is associated with an integer-valued variable $s$ whose adjustment corresponds to a charge-hop update. 
Both sets of variables are subject to PBCs.

Component-wise, we now define the field
\begin{align}
\label{ExtendedSpinDifference}
[ \mathbf{\Delta} \theta ]_i\left( \mathbf{x}+ \frac{a}{2}\mathbf{e}_i \right) :=\frac{ \varphi(\mathbf{x}\!+\!a\mathbf{e}_i)\!-\! \varphi(\mathbf{x})\! +\! q s(\mathbf{x}\!+\!a\mathbf{e}_i,\mathbf{x})}{a} ,
\end{align}
and identify
\begin{align}
\label{MaggsFieldIdentify}
\mathbf{E}(\mathbf{x})\equiv \frac{1}{\epsilon_0}
\left( \begin{array}{c}
[ \mathbf{\Delta}\theta ]_y(\mathbf{x}+\frac{a}{2}\mathbf{e}_x)  \\
\\
-[ \mathbf{\Delta}\theta ]_x(\mathbf{x}+\frac{a}{2}\mathbf{e}_y)  \end{array} \right).
\end{align}
The ${\bf x}$ coordinates in Eq. (\ref{ExtendedSpinDifference}) are in $D'$; the ${\bf x}$ coordinates in Eq. (\ref{MaggsFieldIdentify}) are in $D$.

\begin{figure}[h!tb]
\includegraphics[width=\linewidth]{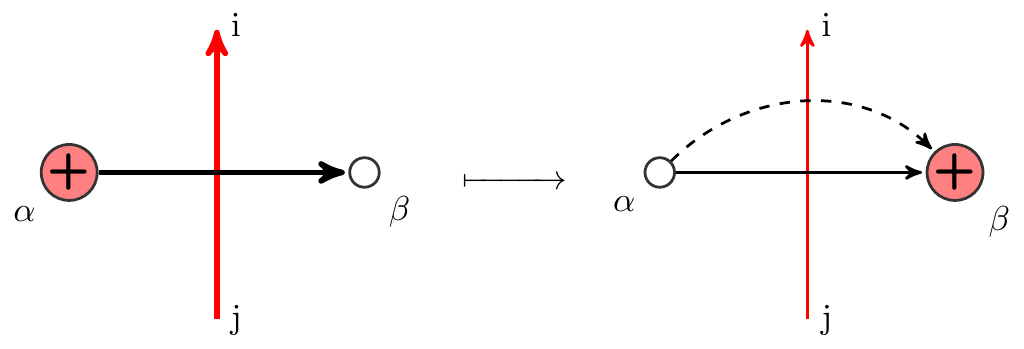}
\caption{A charge-hop update in the positive $x$ direction: The $s_{ij}$ variable (red arrow) has its value decreased by an amount $q$. The value of the electric field flux $E_{\alpha \beta}$ (black arrow) flowing from site $\alpha$ to site $\beta$ then decreases by $q/ \epsilon_0$, corresponding to a charge-hop update. Red circles represent positive charges; white circles represent empty charge sites.} \label{nchargeMR}
\end{figure}

A charge hop in the positive $x/y$ direction corresponds to a decrease/increase in the relevant $s$ variable by an amount $q$, as shown in Fig. \ref{nchargeMR} (where $s_{ij}$ represents the $s$ variable between sites $i$ and $j$ of the conjugate lattice). 

\begin{figure}[h!tb]
\includegraphics[width=\linewidth]{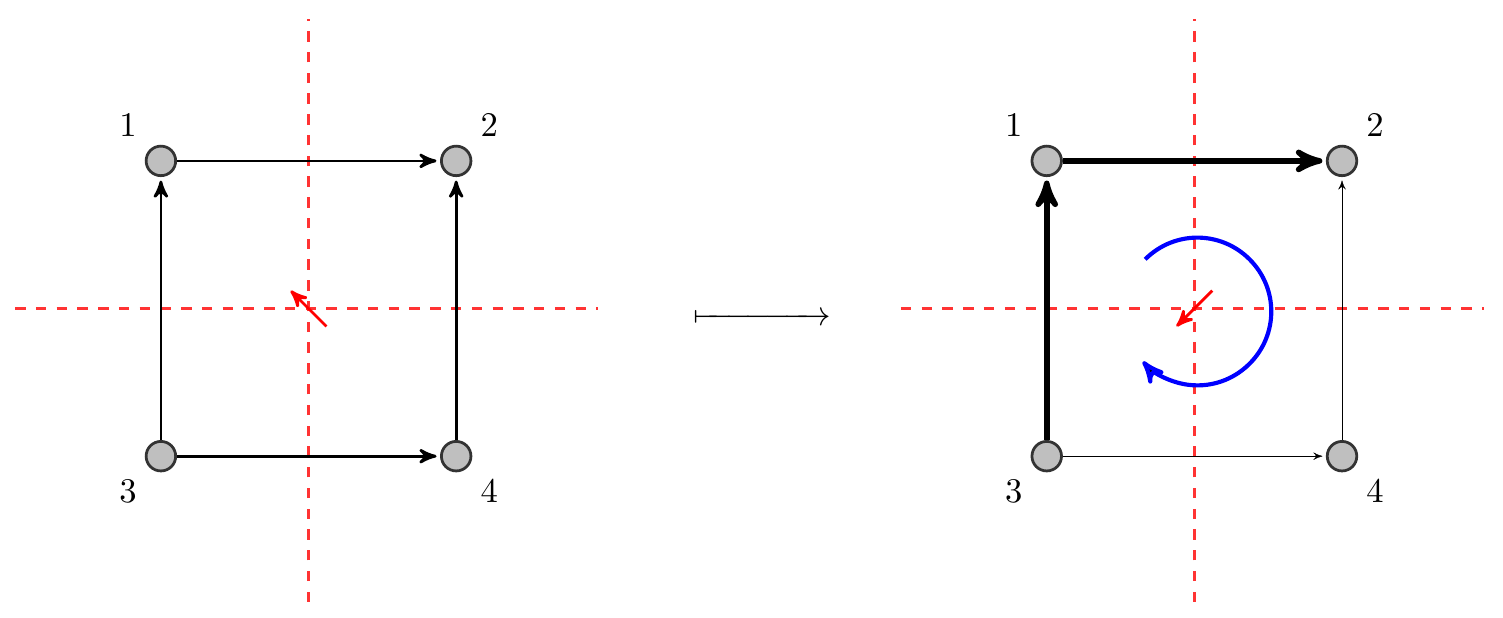}
\caption{An update of the rotational degrees of freedom of the electric field: The value of the $\varphi$ variable at the centre of a randomly chosen lattice plaquette decreases by an amount $\Delta$. This rotates the electric flux by an amount $\Delta /\epsilon_0$ around the plaquette, leaving Gauss' law satisfied. Red arrows represent $\varphi$ variables, black arrows represent the electric field, dashed red lines represent the conjugate lattice $D'$, the blue arrow represents the direction of the field rotation and grey circles represent sites of arbitrary charge.} \label{rotspinMR}
\end{figure}

Fig. \ref{rotspinMR} depicts the microscopic-variable representation of the auxiliary-field updates, with an alteration of a particular $\varphi$ variable rotating the field around its surrounding plaquette. In the figure, the $\varphi$ variables are represented by spin-like arrows in order to emphasize the rotation of the electric field. 

With the internal energy of the electric fields corresponding to a given charge and auxiliary-field configuration given by $U_0$ in Eq. (\ref{u0}), it is possible to write the partition function in the microscopic-variable representation. For ease of manipulation, we allow charge-hop updates to create charge pairs out of the vacuum and include the possibility of all integer-valued multiples of the elementary charge. Combining Eqs. (\ref{u0}) and (\ref{MaggsFieldIdentify}), the partition function in the microscopic-variable representation is given by
\begin{align}
Z=\sum_{\{ s\} }\int \mathcal{D}\varphi &\exp \left[ -\frac{\beta}{2\epsilon_0}\sum_{\langle \mathbf{x},\mathbf{x}'\rangle}\! |\varphi(\mathbf{x})\!-\!\varphi(\mathbf{x}')\!+\!qs(\mathbf{x},\mathbf{x}')|^2 \right]\nonumber\\
\times &\exp \left( -\beta U_{\text{Core}} \right) ,
\end{align}
where
\begin{align}
\int \mathcal{D}\varphi := \prod_{\mathbf{x}\in D'} \left[ \int_{-q/2}^{q/2}d\varphi(\mathbf{x}) \right],
\end{align}
and $\sum_{\{ s\} }:=\sum_{\{ s(\mathbf{x},\mathbf{x}')\in \mathbb{Z}\} }$. Here, the grand-canonical energy of the Coulombic system $U=U_0+U_{\text{Core}}$ is used.

This representation reproduces Gauss' law:
\begin{align}
\sum_{\mathbf{x}\in \partial \Gamma}\mathbf{\Delta}\theta (\mathbf{x})\cdot \mathbf{l}(\mathbf{x})=Q_{\Gamma},
\end{align}
where $Q_{\Gamma}$ is the charge enclosed within some subset of the lattice $\Gamma \subseteq D$, $\partial \Gamma \subset D'$ is the boundary enclosing $\Gamma$, and $\mathbf{l}$ traces an anticlockwise path along $\partial \Gamma$ and has dimensions of length. This equation results from the $ \varphi$ variables cancelling and the $s$ variables being integer valued. It follows that
\begin{align}
\sum_{\mathbf{x}\in \partial \Gamma}\mathbf{\Delta}\theta (\mathbf{x})\cdot \mathbf{l}(\mathbf{x})=&a^2\sum_{\mathbf{x}\in \Gamma}\epsilon_0 \boldsymbol{\hat{\nabla}}\cdot \mathbf{E}(\mathbf{x})  \\
\Rightarrow \boldsymbol{\hat{\nabla}}\cdot \mathbf{E}(\mathbf{x}) =& \rho(\mathbf{x})/\epsilon_0,
\end{align}
recovering Eq. (\ref{gauss}), as required.

The constraints imposed upon the electric field (Gauss' law and the form of the harmonic mode, the latter of which is derived in detail in Appendix \ref{Polarization}) are combined with the grand-canonical energy of the system to write the partition function in terms of the electric field. 
We define the set $X:=q\mathbb{Z}/a^2$, such that the partition function is given by
\begin{align}
Z=&\left| {\bf J} \right| \sum_{\{ \rho(\mathbf{x})\in X\} }\sum_{\mathbf{w}_0\in \mathbb{Z}^2}\int \mathcal{D}\mathbf{E} \exp \left[ -\frac{\beta \epsilon_0 a^2}{2}\sum_{\mathbf{x}\in D}|\mathbf{E}(\mathbf{x})|^2 \right] \nonumber\\
&\times \exp \left( -\beta U_{\text{Core}}\right) \prod_{\mathbf{x}\in D} \left[ \delta\left(\boldsymbol{\hat{\nabla}}\cdot \mathbf{E}(\mathbf{x})-\rho(\mathbf{x})/\epsilon_0\right) \right] \nonumber\\
&\times \delta \left( \sum_{\mathbf{x}\in D}\mathbf{E}(\mathbf{x})+\left( \frac{N}{\epsilon_0}\mathbf{P}-\frac{Lq}{\epsilon_0 a^2}\mathbf{w}_0\right) \right) , 
\end{align}
where the functional integral
\begin{align}
\int \mathcal{D}\mathbf{F}:=\prod_{\mathbf{x}\in D}\left[ \int_{\mathbb{R}} d F_x(\mathbf{x}+a\mathbf{e}_x/2) \int_{\mathbb{R}} d F_y(\mathbf{x}+a\mathbf{e}_y/2) \right]
\end{align}
for any vector field $\mathbf{F}$, and $\left| {\bf J} \right|$ is the Jacobian determinant.

This partition function may be separated into two components by defining the new rotational field
\begin{align}
\label{EhatEtilde}
\mathbf{\tilde{e}}(\mathbf{x}):=\mathbf{E}(\mathbf{x})+\boldsymbol{\tilde{\nabla}}\phi (\mathbf{x})-\mathbf{\bar{E}}.
\end{align}
The partition function is then given by
\begin{align}
Z=Z_{\text{Coul}}\,  Z_{\text{Rot}},
\end{align}
where
\begin{align}
Z_{\text{Coul}}:=\sum_{\{ \boldsymbol{\nabla}^2\phi(\mathbf{x})\in Y\} }&\exp \left[ -\frac{\beta \epsilon_0 a^2}{2}\sum_{\mathbf{x}\in D}|\boldsymbol{\tilde{\nabla}} \phi(\mathbf{x})|^2 \right] \nonumber\\
\times \sum_{\mathbf{w}_0\in \mathbb{Z}^2}&\exp\left( -\frac{\beta }{2\epsilon_0}|L\mathbf{P}-q\mathbf{w}_0|^2 \right) \nonumber\\
\times & \exp \left( -\beta U_{\text{Core}} \right),
\end{align}
and
\begin{align}
Z_{\text{Rot}} :=& \left| {\bf J} \right| \int \mathcal{D}\mathbf{\tilde{e}} \exp \left[ -\frac{\beta \epsilon_0 a^2}{2}\sum_{\mathbf{x}\in D}|\mathbf{\tilde{e}}(\mathbf{x})|^2 \right] \nonumber\\
&\times \prod_{\mathbf{x}\in D} \left[ \delta \left(\boldsymbol{\hat{\nabla}}\cdot \mathbf{\tilde{e}}(\mathbf{x})\right) \right] \delta \left( \sum_{\mathbf{x}\in D}\mathbf{\tilde{e}}(\mathbf{x})\right)  
\end{align}
are the Coulombic and auxiliary-field components of the partition function, respectively. Here $Y:=q\mathbb{Z}/\epsilon_0 a^2$ and we have used the fact that all coupling terms in the grand-canonical energy sum to zero. The MR algorithm therefore reproduces Coulombic physics since the separation of the auxiliary-field partition function from the Coulombic partition function ensures that the charge-charge correlations are independent of the auxiliary field.

The lattice Green's function $G(\mathbf{x},\mathbf{x}')$ between two charge-lattice sites $\mathbf{x}$ and $\mathbf{x}'$ is defined such that
\begin{align}\label{lattice_Greens_Laplacian}
\boldsymbol{\nabla}_{\mathbf{x}}^2G(\mathbf{x},\mathbf{x}') = -\delta_{\mathbf{x},\mathbf{x}'},
\end{align}
where the subscript $\mathbf{x}$ denotes with respect to which coordinate system the lattice Laplacian is applied.

We define the $\mathbf{k}$-space lattice Green's function $\tilde{G}$,
\begin{align}\label{G_tilde}
\tilde{G}_{\mathbf{x}'}(\mathbf{k}) := \sum_{\mathbf{x}\in D}e^{-i \mathbf{k}\cdot \mathbf{x}}G(\mathbf{x},\mathbf{x}'),
\end{align}
and the set $\sum_{\mathbf{k}\in B}:= \prod_{i \in \{ x,y\}}\left[ \sum_{k_{i}\in B_{i}}\right]$, where $B_{i} := \{ 0,\pm \frac{2\pi}{N_{i}a},\pm 2\frac{2\pi}{N_{i}a},\cdots ,\pm (\frac{N_{i}}{2}-1)\frac{2\pi}{N_{i}a} ,\frac{N_{i}}{2}\frac{2\pi}{N_{i}a}  \}$ is the set of $\mathbf{k}$-space values in the $i$ direction, and $N_i:=\sqrt{N}$. Combining Eqs. (\ref{lattice_Greens_Laplacian}) and (\ref{G_tilde}), it then follows that
\begin{align}
\sum_{\mathbf{k}\in B}e^{i \mathbf{k}\cdot (\mathbf{x}-\mathbf{x}')}=2\! \sum_{\mathbf{k}\in B}&e^{i \mathbf{k}\cdot \mathbf{x}}\left[ 2-\cos(k_xa)-\cos(k_ya)\right] \nonumber\\
&\times \tilde{G}_{\mathbf{x}'}\!(\mathbf{k}).
\end{align}
This is solved by
\begin{align}
\label{cases}
\tilde{G}_{\mathbf{x}'}(\mathbf{k})=\frac{e^{-i \mathbf{k}\cdot \mathbf{x}'}}{2\left[ 2\!-\!\cos(k_xa)\!-\!\cos(k_ya) \right]} \, \forall \mathbf{k} \ne \mathbf{0},
\end{align}
where the $\mathbf{k}=\mathbf{0}$ part of the lattice Green's function is set to zero since the harmonic component of $\mathbf{E}$ is attributed to $\mathbf{\bar{E}}$. It follows that
\begin{align}
\label{corrected_greens}
G(\mathbf{x},\mathbf{x}')= \frac{1}{2N} \sum_{\mathbf{k}\ne \mathbf{0}}\frac{e^{i \mathbf{k}\cdot (\mathbf{x}-\mathbf{x}')}}{ 2-\cos(k_xa)-\cos(k_ya)}.
\end{align}

The internal energy of the Poisson component of the electric field is given by
\begin{align}
U_{\text{Poisson}} :=& \frac{\epsilon_0 a^2}{2}\sum_{\mathbf{x}\in D} | \boldsymbol{\tilde{\nabla}} \phi(\mathbf{x})|^2 \\ 
=& - \frac{\epsilon_0 a^2}{2}\sum_{\mathbf{x}\in D} \phi(\mathbf{x})  \boldsymbol{\nabla}^2 \phi(\mathbf{x}) \\
=& \frac{a^4}{2\epsilon_0} \sum_{\mathbf{x}_i,\mathbf{x}_j\in D}\rho(\mathbf{x}_i)G(\mathbf{x}_i,\mathbf{x}_j) \rho(\mathbf{x}_j) ,
\end{align}
hence, the Coulombic partition function can be written as
\begin{align}
Z_{\text{Coul}}=\sum_{\{ \rho(\mathbf{x})\in X\} }&\exp \left[ -\frac{\beta a^4}{2\epsilon_0}\sum_{\mathbf{x}_i,\mathbf{x}_j\in D}\rho(\mathbf{x}_i)G(\mathbf{x}_i,\mathbf{x}_j)\rho(\mathbf{x}_j) \right] \nonumber\\
&\times \sum_{\mathbf{w}_0\in \mathbb{Z}^2}\exp\left( -\frac{\beta }{2\epsilon_0}|L\mathbf{P}-q\mathbf{w}_0|^2 \right)\nonumber\\
&\times \delta \left( \sum_{{\bf x}}\rho ({\bf x}) \right) \exp \left( -\beta U_{\text{Core}} \right),
\end{align}
where the $\delta$ function enforces charge neutrality in the Green's function representation.

\section{Polarization}\label{Polarization}

We consider the sum of each component of the electric field over the entire lattice in order to analyse the harmonic mode. The sum of the $x/y$-component is split into separate sums over all $x/y$-components that enter a particular strip of plaquettes of width $a$ that wrap around the torus in the $y/x$ direction.  Each component of the harmonic mode $\bar{E}_{x/y}$ is then expressed in terms of the charge enclosed along each of the strips of plaquettes:
\begin{align}
L^2 \bar{E}_x=&a^2\sum_{\mathbf{x}\in D}E_x\left( \mathbf{x}+\frac{a}{2}\mathbf{e}_x \right) \\
=& a\sum_{x=0}^{L-2a} (x\!+\! a)\sum_{y=0}^{L-a} \left[ E_x\! \left( x\!+\!\frac{a}{2}, y \right) \! -\! E_x\! \left( x\!+\!\frac{3a}{2}, y\right) \right] \nonumber\\
&+ La\sum_{y=0}^{L-a} \left[ E_x \left( L-\frac{a}{2}, y\right) - E_x\left( \frac{a}{2}, y\right) \right] \nonumber\\
&+ La\sum_{y=0}^{L-a}E_x\left( \frac{a}{2}, y\right) \\
\label{chargedensity}
=& -\frac{a^2}{\epsilon_0} \sum_{x=a}^{L} x \sum_{y=a}^{L} \rho(\mathbf{x}) + La\sum_{y=a}^{L}E_x \left( \frac{a}{2}, y \right) ,
\end{align}
which follows from applying Gauss' law to each strip of plaquettes that wrap around the torus in the $y$ direction. The same argument holds for the $y$ component, hence, the harmonic mode is given by
\begin{align}
\mathbf{\bar{E}}=-\frac{1}{\epsilon_0}\mathbf{P}+\frac{q}{L\epsilon_0}\mathbf{w}_0,
\end{align}
where $\mathbf{P}:= \sum_{\mathbf{x}\in D}\mathbf{x} \rho(\mathbf{x}) /N$ is the origin-dependent polarization vector of the system and $w_{0,x}:=\epsilon_0 a\sum_{y=a}^LE_x(a/2,y)/q$ is the $x$ component of the origin-dependent winding field, with the $y$ component defined analogously. Here, $\mathbf{P}$ and $\mathbf{w}_0$ are measured from a specific origin. Note that the above applies to systems composed of either single- or multi-valued charges.

We have thus shown that $\mathbf{\bar{E}}$, which is origin-independent, is given by the sum of two origin-dependent terms.  One of these is attributed to the polarization of the system, while the other describes the winding of charges around the torus given that the polarization is measured with respect to the chosen origin.

Restricting our attention to the gas of elementary charges, we now devise an origin-independent measure of the topological sector of the system. 
First, we note that adding $\omega$ windings to either component of the harmonic mode $\mathbf{\bar{E}}$ corresponds to
 \begin{align}
 \bar{E}_{x/y}\mapsto  \bar{E}_{x/y} + \frac{q}{L\epsilon_0}\omega,
 \end{align}
and that this results in a change in the grand-canonical energy of the system given by
\begin{align}
\label{dipdef}
\Delta U= \frac{Lq}{2}\omega \left( \frac{q}{L\epsilon_0}\omega + 2\bar{E}_{x/y} \right) .
\end{align}
Hence, given an arbitrary charge distribution, the lowest-energy harmonic mode that describes the charge distribution is an element of the set 
in Eq. (\ref{dipset}). 
We therefore define a convention in which the harmonic mode is given by Eq. (\ref{approach2}), where the polarization component of the harmonic mode is an element of the set in Eq. (\ref{dipset}) and the winding component of the harmonic mode is given by Eq. (\ref{Ebarw}).

\section{Simulation details}
\label{sim_details}

The system was simulated using the MR algorithm on an $L\times L$ lattice of lattice spacing $a=1$. One charge-hop sweep corresponded to picking a charge site at random, picking the $x$ or $y$ direction at random, then proposing a charge hop in the positive or negative direction (at random), repeating this $2N$ times (replacing each site / field bond after each proposal). One auxiliary-field sweep corresponded to picking a charge site at random and proposing a field rotation around the site, repeating this $N$ times. One global sweep corresponded to proposing a winding update in the positive or negative (at random) $x$ and $y$ directions. For all simulations, we performed five auxiliary-field sweeps per charge-hop sweep, and, for those simulations that also employed the global update, we performed one global update per charge-hop sweep. One charge-hop sweep corresponds to one Monte Carlo time step.

The data sets in Sections \ref{ergodicity} and \ref{FSS} were averaged over multiple runs of $10^6$ charge-hop sweeps per lattice site. The data set in Fig. \ref{figsuscquotient} was averaged over $608$ and $446$ runs between $T=1.15$ and $1.45$ with the global update off and on, respectively, over $384$ runs between $T=1.5$ and $1.6$, and over $256$ runs between $T=1.65$ and $1.75$.

The $L=8$ data set in Fig. \ref{figfullsusc} was averaged over $128$ ($T=0.1-1.1$), $256$ ($T=1.15-1.39;T=1.41-1.44;T=1.46- 1.49$), $768$ ($T=1.4;T=1.45; T=1.5-1.75$), and $256$ ($T=1.8-2.5$) runs; the $L=16$ data set was averaged over $128$ ($T=0.1-1.1$) 
and $256$ ($T=1.15-2.5$) runs; the $L=32$ data set was averaged over $128$ ($T=0.1-1.1$), $256$ ($T=1.15-2.0$), and $128$ ($T=2.0-2.5$) runs; the $L=64$ data set was averaged over $128$ ($T=0.1-1.1$), $448$ ($T=1.15-1.45$), $384$ ($T=1.5-1.6$), $256$ ($T=1.65-2.0$), and $128$ ($T=2.05-2.5$) runs.

We also simulated the $L=10$, $L=20$, and $L=40$ systems over small temperature ranges to calculate additional crossover points for Fig. \ref{figtempextrap}: all data sets were averaged over $512$ runs.

\bibliography{Bibliography_new}{}

\end{document}